\documentstyle[psfig]{article} 

\def\1ad{\mbox{\normalsize $^1$}}
\def\2ad{\mbox{\normalsize $^2$}}
\def\3ad{\mbox{\normalsize $^3$}}
\def\4ad{\mbox{\normalsize $^4$}}
\def\5ad{\mbox{\normalsize $^5$}}
\def\6ad{\mbox{\normalsize $^6$}}
\def\7ad{\mbox{\normalsize $^7$}}
\def\8ad{\mbox{\normalsize $^8$}}
\def\makefront{\vspace*{1cm}\begin{center}
\def\newtitleline{\\ \vskip 5pt}
{\Large\bf\titleline}\\
\vskip 1truecm
{\large\bf\authors}\\
\vskip 5truemm
\addresses
\end{center}
\vskip 1truecm
{\bf Abstract:}
\abstracttext
\vskip 1truecm}
\newcommand{\tr}[1]{\:{\rm Tr}\,#1}

\newcommand{\rf}[1]{(\ref{#1})}
\newcommand{\eq}[1]{Eq.~(\ref{#1})}

\def\be{\begin{equation}}
\def\ee{\end{equation}}
\def\bea{\begin{eqnarray}}
\def\eea{\end{eqnarray}}

\begin{document}
\begin{flushright} 
{\large ITEP--TH--9/99 \\
hep-th/9903030}
\end{flushright} 
 
\def\titleline{Formulation of Matrix Theory 
at Finite Temperature\footnote{
Based on the paper~\cite{ams98} written in collaboration with
Jan Ambj{\o}rn and Gordon Semenoff.}} 
\def\authors{Yuri Makeenko}
\def\addresses{Institute of Theoretical and Experimental 
Physics\\ B. Cheremushkinskaya 25, 
117218 Moscow, Russia\\ \tt makeenko@itep.ru }

\def\abstracttext{
The interaction between static D0-branes at finite temperature is
considered in the matrix theory and the superstring theory.
The results agree in both cases to the leading order in the 
supersymmetry violation by temperature, where the one-loop approximation 
is reliable.
The effective static potential is short-ranged and attractive. 

Talk at the 32nd 
International Symposium Ahrenshoop on the Theory of
Elementary Particles, 
Buckow, Germany, September 1--5, 1998. }

\large
\makefront

\section{Introduction}

The matrix theory~\cite{bfss} is the large-$N$ limit of the 
10-dimensional supersymmetric Yang--Mills theory
dimensionally reduced to 0~spatial dimensions.
When the coupling constant $g^2_{\rm YM}$ is large, 
the matrix theory describes 11-dimensional M-theory while
the limit of small $g^2_{\rm YM}$ is associated with 10-dimensional
\mbox{IIA superstring.} The~matrix theory correctly reproduces
properties of D-branes in the superstring theory 
including their interactions to the leading order in 
violation of supersymmetry, e.g.\ at small velocities or
large separations between D-branes or weak magnetic fields
living on D-branes.

In this talk I consider the formulation of the matrix theory at finite 
temperature given by an Euclidean path integral with
boundary conditions along the compactified ``time''  
which are periodic for the Yang--Mills fields
and antiperiodic for fermionic superpartners. 
I present the result of the computation of
the effective potential between static D0-branes in the one-loop
approximation and show that it agrees  
with an analogous computation in superstring theory, where an
integration is to be performed over the non-trivial holonomies
of the temporal components of Abelian gauge fields living 
on the D0-branes. 
This agreement is to the leading order in the supersymmetry violation 
by temperature, where the one-loop approximation is reliable,
thus providing one more argument supporting the validity of the matrix theory. 
The computed effective static potential which is short-ranged and attractive
has consequences for thermal properties of D0-branes. 

\section{Matrix theory at finite temperature}

The matrix theory~\cite{bfss} is formulated
by the reduction of ten dimensional supersymmetric 
Yang-Mills theory 
\begin{equation}
S_{\rm YM}[A,\theta]=\frac{1}{g_{YM}^2}\int d\tau\, {\rm Tr}\left(
\frac{1}{4}F_{\mu\nu}^2+\frac{i}{2}\theta\gamma_\mu D_\mu\theta\right)
\end{equation}
to one temporal and
zero spatial dimensions: $A_\mu=A_\mu(\tau)$, 
$\theta=\theta(\tau)$. 

The thermal partition function is given by the Euclidean path integral
\begin{equation}
Z_{\rm YM}=\int[dA(\tau)][d\theta(\tau)]e^{ -S_{\rm YM}[A,\theta]},
\label{thermal}
\end{equation}
where the time-coordinate is
periodic. The bosonic and fermionic coordinates have, respectively,  
periodic and
antiperiodic boundary conditions
\begin{eqnarray}
A_\mu(\tau+\beta)= A_\mu(\tau),  ~~~
\theta(\tau+\beta)=-\theta(\tau),~~~
\beta=1/T ,
\label{b.c}
\end{eqnarray}
where $T$ is the temperature.  Gauge
fixing involves introducing ghost fields
which  have periodic boundary conditions.

The representation (\ref{thermal}) of the  thermal partition 
function can be derived in the standard way starting from the known
Hamiltonian of the matrix theory~\cite{bfss} and representing
the  thermal partition function 
\begin{equation}
Z_{\rm YM}= {\rm Tr}\, e^{-\beta H}  
\end{equation}
via the path integral. The trace is calculated over all states
 obeying Gauss's law which is taken care by the integration
over $A_0$ in (\ref{thermal}). This representation of the  matrix theory
at finite temperature have been discussed 
in Refs.~\cite{OZ98,MOP98,Sat98}.

The diagonal components of the gauge fields,
$\vec{a}^\alpha\equiv \vec{A}^{\alpha\alpha}$, are interpreted 
in the matrix theory as the
positions of the $\alpha$-th D0-brane and should be
treated as collective variables. Static configurations play a special
role since they satisfy classical equations of motion 
with the periodic boundary conditions and dominate
the path integral as $g^2_{\rm YM}\rightarrow 0$. There
are no such static zero modes for fermionic components since they would not
satisfy the antiperiodic boundary conditions.
This is an important
difference from the  zero temperature case and a manifestation of the fact
that supersymmetry is explicitly broken by non-zero temperature.

An effective action for these coordinates is constructed 
by integrating the
off-diagonal components of the 
gauge fields, the fermionic variables and the ghosts:
\begin{equation}
S_{\rm eff}[ \vec a^\alpha]\equiv -\ln \int
\prod_{\beta\neq\alpha}[d a^\alpha_0]
[dA^{\alpha\beta}_\mu][d\theta][d{\rm ghost}]e^
{-S_{\rm YM}-S_{\rm gf}-S_{\rm gh}}.
\end{equation}
Generally, this integration can only be done in the a simultaneous
loop expansion and expansion in the number of derivatives of the
coordinates $\vec a^\alpha$.  Such an expansion is accurate in the
limit where $\left| \vec a^\alpha-\vec a^\beta\right|$ are large for
each pair of D0-branes and where their velocities are small.   
The remaining dynamical problem
then defines the statistical mechanics of a gas of D0-branes:
\begin{equation}
Z_{\rm YM}=\int\prod_{\tau,\alpha}
[d \vec a^\alpha(\tau)]e^{-S_{\rm eff}[\vec a^\alpha]}.
\label{effec}
\end{equation}

\section{One-loop computation in matrix theory}

The computation of the effective action $S_{\rm eff}$ 
for the interaction between static D0-branes at one loop
is standard for the matrix theory.

The gauge field decomposed into diagonal part which satisfies
the classical equation of motion and fluctuating
off-diagonal part:
\begin{equation}
A_\mu^{\alpha\beta}=a_\mu^{\alpha}\delta^{\alpha\beta}+g_{\rm YM}
\bar A^{\alpha\beta}_\mu .
\end{equation}
The gauge is fixed by 
\begin{equation}
D_\mu^{\alpha\beta}\bar A_\mu^{\alpha\beta}=0,
\end{equation}
where 
\begin{equation}
D_0^{\alpha\beta}=\partial_0
-i\left(a^\alpha_\mu-a^\beta_\mu\right),
~~~ \vec D^{\alpha\beta}=-i\left(\vec a^\alpha-\vec a^\beta\right).
\end{equation}
This adds the Fadeev-Popov ghosts to the action
\begin{equation}
S_{\rm gh}=\int\sum_{\alpha\beta} \left\{\bar c^{\alpha\beta}\left( 
-D^{\alpha\beta}_\mu\right)^2c^{\beta\alpha}+i
g_{\rm YM}\bar c^{\beta\alpha}D_\mu^{\alpha\beta}
\left[\bar A_\mu,c\right]\right\}.
\end{equation}

There is a residual Abelian gauge invariance
\begin{equation}
\bar A^{\alpha\beta}_\mu\rightarrow \bar A^{\alpha\beta}_\mu 
e^{i(\chi^\alpha -\chi^\beta)},~~~~
a_\mu^\alpha\rightarrow a_\mu^\alpha+\partial_\mu\chi^\alpha,
\end{equation}
which can be used to make $a_0^{\alpha}$ independent
on the compactified time-variable
($\partial_0 a^{\alpha}_0=0 $).
In contrast to the zero-temperature case, $a_0^{\alpha}$'s
can not be completely removed bacause of the existence of
the nontrivial holonomy 
\begin{equation}
{\rm P} e^{i \int_0^\beta d\tau A_0(\tau)}= 
\Omega^\dagger\;{\rm diag}\; \left(
e^{i\beta a_0^1},\ldots,e^{i\beta a_0^{\rm N}}\right) \Omega ,
\end{equation}
which is
known as the Polyakov loop winding around the compact Euclidean time,
whose trace is gauge invariant.
Due to periodicity, it is chosen
$-\pi/\beta < a_0^\alpha\leq\pi/\beta$.

Expanding the action to the 
 quadratic order in $\bar A, c,\bar c, \theta$ and doing
the Gaussian integration, it is obtained in the standard way
\begin{equation}
S_{\rm eff}=8\sum_{\alpha <\beta}\left\{{\rm
Tr}_B\ln\left( -(D^{\alpha\beta}_\mu)^2 \right) -{\rm Tr}_F\ln\left(
-(D^{\alpha\beta}_\mu)^2 \right)\right\},
\end{equation}
where the subscript $B$ denotes contributions from the gauge fields
and ghosts, whereas $F$ denotes those from the adjoint fermions.  The
determinants should be evaluated with periodic boundary conditions for
bosons and antiperiodic boundary conditions for fermions.  

The boundary conditions are taken into account by proper Matsubara 
frequencies, so that
\begin{equation}
e^{-S_{\rm eff}}=\beta^{\rm N}
\int_{-\pi/\beta}^{\pi/\beta} \prod_{\gamma>\alpha}\frac{da^\alpha_0}{2\pi} 
\prod_{n=-\infty}^\infty
\left( \frac{ \left(\frac{2\pi n}{\beta}
+\frac{\pi}{\beta}+a_0^\alpha-a_0^\gamma\right)^2+\vert\vec 
a^\alpha-\vec a^\gamma\vert^2}{\left(  
\frac{2\pi n}{\beta}+a_0^\alpha-a_0^\gamma\right)^2 +\vert\vec 
a^\alpha-\vec a^\gamma\vert^2}
\right)^8.
\end{equation}
Using the formula
\begin{equation}
\prod_{n=-\infty}^\infty
\left( \frac{2\pi n}{\beta}+\omega\right)=\sin\left( \frac{\beta\omega}
{2}\right),
\label{21}
\end{equation}
we obtain finally
\begin{equation}
e^{-S_{\rm eff}}=\beta^{\rm N}
\int_{-\pi/\beta}^{\pi/\beta} \prod_{\gamma>\alpha}
 \frac{da^\alpha_0}{2\pi} 
\left(  \frac{ \cosh \beta\vert\vec a^\alpha-\vec a^\gamma\vert+
\cos \beta \left(a_0^\alpha- a_0^\gamma\right)}
{\cosh\beta\vert\vec a^\alpha-\vec a^\gamma\vert - 
\cos\beta \left( a_0^\alpha-a_0^\gamma\right) }\right)^8.
\label{23}
\end{equation}
The integration over
 the temporal components  $a_0^\alpha$   implements the projection 
onto the gauge invariant eigenstates of the matrix theory Hamiltonian.


If both bosons and fermions had periodic
boundary conditions the determinants would cancel
because of supersymmetry.  This
would give the well-known result that the lowest energy state is a BPS
state whose energy does not depend on the relative separation of the
D0-branes.
  
\section{Comparison with superstring theory}

\subsection{Open string language}

The starting point is the thermal partition function
of the single open superstring:  
\be
Z_{\rm 1str} \equiv -\beta {\cal F} = \tr e^{-\beta H},
\label{1string}
\ee
where $H$ is the superstring Hamiltonian and the trace is
over all physical (GSO projected) superstring states. 

Since the ends of the open string end on two D0-branes separated
by the distance $L$, the superstring has
the Neumann boundary condition along the temporal direction and
the Dirichlet boundary conditions along the nine spatial directions.
The corresponding superstring spectrum is given by 
\begin{equation}
\sqrt{\alpha'}E_N=\sqrt{\frac{L^2}{4\pi^2\alpha'}+N},
\label{Esuperstring}
\end{equation}
where $N$ are eigenvalues of the oscillator number operator. 

Knowing the spectrum \rf{Esuperstring}, the 
thermal partition function of the string gas can be immediately
written as 
\begin{equation}
Z_{\rm str}(\beta,L,\nu)= e^{-\beta {\cal F}}= \prod_{N=0}^\infty \left|
\frac{1+e^{-\beta E_N+i\pi \nu}}{1-e^{-\beta E_N+i\pi \nu}}
\right|^{2d_N},
\label{Zsuperstring}
\end{equation}
where $d_N$ stands for the degeneracy of  
either bosonic of fermionic superstring states at level $N$:
\begin{equation}
8\prod_{n=1}^\infty\left( \frac{1+e^{-nl}}{1-e^{-nl}}\right)^8
=\sum_{N=0}^{\infty} d_N e^{-Nl} .
\label{degeneracy}
\end{equation}
For the lowest levels, $d_0=8$ and $E_0=L/2\pi\alpha'$.
The factor of 2 in the exponent $2d_N$ in (\ref{Zsuperstring}) 
is the famous one by Polchinski~\cite{Pol95} and 
is due to the interchange of the superstring ends. 
It is crucial in  providing the agreement with the matrix theory 
computation.

The physical meaning of \eq{Zsuperstring} is obvious:
the partition function is equal to the ratio of the Fermi and
Bose distributions with the power being 
(twice) the degeneracy of the states. 

\begin{figure} 
\hspace{1.5in}
\psfig{figure=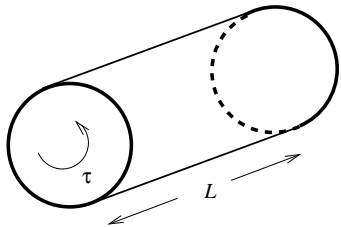,height=1.5in} 
\caption[x]{\footnotesize Space-time cylinder. 
The bold lines represent world-lines of the two D0-branes which go
along the periodic temporal direction and are separated
by the distance $L$. 
They bound the string world-sheet which can wind around the
cylinder.} 
\label{f:annulus} 
\end{figure}

Equation~\rf{Zsuperstring} is derived in Ref.~\cite{green}
(in Ref.~\cite{VM96} for Dp-branes)
by calculating the annulus diagram for the open superstring
in compactified Euclidean time of the circumference $\beta$.
The parameter $\nu$ has the meaning of the constant U(1) gauge
field which enters though the quantized temporal momentum
$p^0={2\pi(r-\nu)}/{\beta}$
of the open string whose world-sheet winds aroung the
space-time cylinder as is depicted in Fig.~\ref{f:annulus}.
In the above formula, $r$ is integer in the NS sector 
(associated with space-time bosons) and half-integer
in the R sector (associated with space-time fermions).

In order to compare with the Yang-Mills computation, we identify 
the coordinates of D0-branes with
$\vec q^\alpha= 2\pi\alpha'\vec a^\alpha $,
so that the separation is 
$L=2\pi\alpha' \vert \vec a^1-\vec a^2 \vert$.
Then the integrand in (\ref{23}) coincides for N=2 
with (\ref{Zsuperstring})
truncated to the massless modes ($N=0$) provided 
$\pi\nu=\beta(a_0^1-a_0^2)$.

The truncation of the stringy modes 
is justified for $\beta \gg L$ (or $TL\ll 1$) when
the energy gap $\Delta$ between the first two levels is finite.
From (\ref{Esuperstring}) we get
\begin{equation}
 \Delta =  
\sqrt{ \left(  \frac{L}{2\pi\alpha'}\right)^2+\frac{1}{\alpha'} }-\frac{L}
{2\pi\alpha'}
\label{trunk}
\end{equation}
and the spectrum can be truncated at the first level when
and only when $\beta \Delta \gg 1$.
If the temperature is small, this condition is always satisfied unless
the length $L$ is not too large. 

The integration over $a_0$'s in \eq{23} corresponds to the
integration over $\nu$ in \eq{Zsuperstring}.
This integration comes about in the string theory as follows.
The open-string gauge field $A_0$ interacts with D0-branes 
adding the surface term to the action:
\begin{equation}
S_{\rm int}=\int d q^\mu {A}_\mu=
\int_0^\beta d\tau \left( {A}_0(\tau,\vec q^1)-
{A}_0(\tau,\vec q^2)\right) = \pi \nu
\end{equation}
The matrix theory automatically takes into account the
integration over $A_0$ while in the string theory
calculation of Ref.~\cite{green} the open-string gauge field is fixed.
This intergation over $A_0$ is needed to provide Gauss's law for
the charges at the ends of the open string which are
induced on D-branes. Therefore, the effective potential 
between static D0-branes in the superstring theory is
given by
\begin{equation}
S_{\rm eff}[\vec a^\alpha]= - \ln 
\int_{-1}^{1} d\nu\, Z_{\rm str}(\beta,L,\nu).
\label{Seffsuperstring}
\end{equation} 
This issue will be further discussed in the next Subsection.

\subsection{Closed string language}

At least two issues remain unclear in the open string language.
Firstly, why to exponetiate the single string partition
function to get the string gas and, secondly,
why to integrate over $\nu$ the string gas partition function
(\ref{Zsuperstring}) rather than,
say, the single string partition function \rf{1string}?
This has a natural explanation in the closed string language.

The passage to a closed string is performed  by the
standard modular transformation which converts the annulus diagram
for an open string into a cylinder diagram for a closed string. 
The right-hand side of \eq{1string} can then be represented as~\cite{green}
\begin{equation}
{\cal  F}\left(L,\beta,\nu\right)= \frac{8 \pi^4}{\sqrt{2 \pi \alpha'}} 
\int_0^\infty \frac{ds}{s^{9/2}}\;
e^{s-L^2/2s\alpha'}
\Theta_2 \left(\nu \left \vert \frac{i\beta^2 s}{2\pi^3 \alpha' } 
\right. \right)
\prod_{n=1}^\infty\left( \frac{1-e^{-(2n+1)s}}{1-e^{-2ns}}\right)^8,
\label{viathetas}
\end{equation}
where
\begin{equation} 
\Theta_2 \left(\nu \left \vert iz \right. \right)
=\sum_{q=-\infty}^\infty
e^{ -\pi z(2q+1)^2/4+i\pi(2q+1)\nu}.
\end{equation}
The meaning of \eq{viathetas} is that of the closed-string
propagator, which describes the interaction between D0-branes,
rather than the thermal partition function as for an open string. 

The sum over $q$ in \eq{viathetas} represents the sum over all possible
winding numbers $w=2q+1$ of the closed string around the compact
dimension $X_0$. 
Only odd winding numbers survive since the contribution
of the even ones vanishes due to supersymmetry.
The vanishing of the term with zero winding number is analogous
to that at zero temperature and is due to the cancellation
between the NS-NS and R-R sectors.

When two D0-branes interact, they can exchange several closed
strings, not necessarily one. All such exchanges are of the
same order of magnitude in the string coupling constant
and exponentiate since the closed strings are identical.
This is analogous to the exponentiation of the single-gluon
exchange when the interaction between static quark and antiquark
in the Yang--Mills theory is calculated via the correlator
of two Polyakov loops.
Therefore, \eq{Zsuperstring} naturally emerges in the closed
string language. It is also clear why there is only one $\nu$
for each multi-string term: we have just two interacting D0-branes rather
than a gas of D0-branes. This results in \eq{Seffsuperstring}.

Each of the closed strings mediating the interaction between D0-branes
has its own winding number $w_i$. In the open string language,
this induces on the D-brane
the charge $\sum_i w_i$ with respect to the open-string
gauge field. Such charged states look suspicious since they
are missing at zero temperature
where $X_0$ is not compact and there are no windings along 
the $X_0$ direction, so that the
total charge equals zero at each value of the time $\tau$. 
But the integration over $\nu$ picks up exactly the state with
$\sum_i w_i=0$, i.e.\ which is not charged! In particular, all states 
with a single closed string vanish after the integration over $\nu$.
The leading order contribution to the D0-brane interaction
comes from the state with two closed strings having unit winding
numbers of opposite signs.

It is worth nothing that in the closed string language
$w_i$ is associated with the NS-NS charge of the closed string.
Therefore, the condition $\sum_i w_i=0$ implies the vanishing
of the total NS-NS charge. 

\section{Discussion}

The effective static potential between two D0-branes emerges because 
supersymmetry
is broken by finite temperature. This effect of breaking supersymmetry
is somewhat analogous to the velocity effects at zero temperature
where the matrix theory and superstring computations agree to
the leading order of the velocity expansion~\cite{dkps}.
It is thus shown that the leading term in a low temperature expansion 
is correctly reproduced by the matrix theory.

The effective static potential between D0-branes at one loop
is logarithmic and attractive at short distances.
The singularity 
occurs when the distance between the D0-branes vanishes and the 
SU(N) symmetry which is broken by finite distances is restored.
The integration over the off-diagonal components 
of the gauge field can no longer be treated
in the one-loop approximation! (This issue has been further discussed
recently in Ref.~\cite{BS99}.)
In the superstring theory, the singularity is exactly the same as in the
matrix theory since it is determined only by the massless bosonic modes
in the NS sector. 

The computed partition functions take
into account only thermal fluctuations of superstring stretched
between D0-branes but not the fluctuations of D0-branes themselves.
These separation of the degrees of freedom is justified by the fact
that D0-branes have a mass $1/g_s\sqrt{\alpha'}$ and are very heavy
as $g_s\rightarrow0$. 
To calculate the thermal partition function of D0-branes,
a further path integration over their periodic trajectories $\vec a(\tau)$ 
is to be performed as in (\ref{effec}).  
Classical statistics is not applicable to this problem 
due to the singularity of the 
one-loop effective static potential at small distances
in spite of tha fact that D0-branes are very heavy.
However, this singularity is only in the classical partition function.  
The path integral over the periodic trajectories $\vec a(\tau)$ 
can not diverge since 
the two-body quantum mechanical problem has a well-defined spectrum. 

\vskip0.5cm
\noindent
{\large \bf Acknowledgements}

\smallskip
\noindent
This work is supported in part by the grants 
INTAS 96--0524  and RFFI 97--02--17927.


\end{document}